\documentclass[12pt]{article}
\usepackage{graphicx}
\usepackage{amssymb,epsfig}
\usepackage{lscape,graphics,amsmath}
\usepackage{latexsym,amsfonts}
\usepackage{cite}
\usepackage[english]{babel}

\title{Quantum field-theoretical description of\\ solar neutrino oscillations}
\author{Vadim Egorov and Igor Volobuev}

\begin{document}

\maketitle

\begin{abstract}
A quantum field-theoretical description of neutrino oscillations
in matter in adiabatic approximation is developed. Estimates of
the oscillation coherence lengths have been made for several solar
neutrino production and detection processes. It has been shown
that, within the framework of the developed approach, the results
of solar neutrino detection experiments can be correctly described
in terms of neutrino mass eigenstates without taking oscillations
into account.

\end{abstract}

\section{Introduction}
The phenomenon of neutrino oscillations, which is currently
considered as neutrino changing its flavor depending on the
distance traveled, provides a definitive experimental evidence of
non-zero neutrino masses and gives an explanation of the deficit
of solar neutrinos. Yet the phenomenon cannot be described in the
framework of the standard perturbative S-matrix formalism, because
neutrino oscillation processes occur over finite space and time
intervals, whereas the S-matrix is suitable only for describing
processes that last infinitely long. For this reason neutrino
oscillation processes are usually described either in a quantum
mechanical approach using plane waves
\cite{Pontecorvo:1957cp,Gribov:1968kq, Eliezer:1975ja, Fritzsch:1975rz, Bilenky:1976cw, Bilenky:1976yj,giunti1991,Giunti:2007ry,Bilenky:2010zza,Petcov},
or quantum mechanical and quantum field-theoretical approaches
based on wave packets
\cite{Kayser:1981ye,Giunti:1993se,Grimus:1996av,Beuthe:2001rc,Lobanov:2015esa}.

The quantum mechanical plane-wave description relies on the
concept of neutrino flavor states, which are defined as coherent
superpositions of neutrino mass eigenstates. It is assumed that it
is the neutrino flavor states that are produced in weak
interactions. However, within the plane-wave approximation, the
production of states without a well-defined mass leads to
violation of energy-momentum conservation.  This problem  has been
extensively discussed in the literature
\cite{Giunti:1993se,Grimus:1996av,Beuthe:2001rc,Lobanov:2015esa,Blasone:2019rxl}.
The issue can be resolved by adopting a wave-packet formalism
\cite{Giunti:2007ry}, though at the cost of significantly
increased computational complexity.

The first attempt to describe neutrino oscillations within quantum
field theory was made back in 1982 in paper \cite{Okun1982}. The
approach used the standard perturbative S-matrix formalism, where
the neutrino mass eigenstates produced in a source were assumed to
be virtual particles described by the standard Feynman
propagators, and the oscillations appeared as a result of
interference of the amplitudes corresponding to different mass
eigenstates. The production and detection of neutrinos occurred in
interactions with nuclei, and the matrix elements of the charged
weak hadron currents   of the nuclei were taken to be proportional
to the  delta functions of their positions, while other particles
were described by plane waves. The use of delta functions for
fixing the positions of the nuclei broke the translational
invariance of the theory and resulted in violation of momentum
conservation. Later, the idea was developed in paper
\cite{Giunti:1993se}, where  wave packets were used to describe
the localization of nuclei. However, calculations in the framework
of the wave packet approach turned out to be very complicated
\cite{Naumov:2010um}.

Another approach to describing neutrino oscillations within the
standard perturbative S-matrix formalism was presented in papers
\cite{Pilaftsis:1999,Pilaftsis:2023,Torres:2020,Libanov:2024ukn}.
Essentially, this approach relies on the Bogoliubov procedure
involving the ``switching on" of the interaction \cite{BOSH}. Its
advantage lies in the ability to use the S-matrix formalism in its
original form. However, the introduction of a specific interaction
switching-on function breaks the translational invariance of the
theory and can lead to a violation of energy-momentum
conservation.

A consistent  quantum field-theoretical description of neutrino
oscillations without violation of  energy-momentum conservation
was put forward in papers
\cite{Volobuev:2017izt,Egorov:2017qgk,Egorov:2017vdp}. The central
idea of the approach is to adapt the standard S-matrix formalism
to describing processes of finite duration. Its implementation was
built on the seminal paper by R. Feynman \cite{Feynman:1949zx}.
Specifically, the production and detection processes were treated
as a unified process and the off-shell neutrino mass eigenstates
were described by the standard Feynman propagators. The amplitude
of the total process was first constructed in the coordinate
representation and the transformation to the momentum
representation was performed in a manner tailored to the
experimental situation. Namely, the distance between the
production and detection points was fixed by introducing a delta
function depending on that distance. In practice, this results in
a modification of the Feynman propagators of neutrino mass
eigenstates in momentum space, while all other momentum-space
Feynman rules remain unchanged.

In papers \cite{Egorov:2019vqv,Egorov:2021iig} it was shown that
the approach is capable of explaining the origin of the coherence
length of neutrino oscillations and describing neutrino
oscillations in external magnetic fields. In the present paper we
will show that the approach can also correctly describe neutrino
oscillations in matter and explain the solar neutrino deficit.
Moreover, the obtained results  mean that the only states suitable
for describing neutrino oscillations are the neutrino mass
eigenstates and, paradoxically, neutrino oscillations do not play
any role in explaining the deficit of solar neutrinos.

\section{Modified perturbative formalism}

We will work in the  minimal extension of the Standard Model (SM)
by the right neutrino singlets, where the charged current
interaction Lagrangian of leptons looks like
\begin{equation}\label{L_cc}
\mathcal{L}_{\text{cc}} = - \frac{g }{2\sqrt{2}}\left(
\sum\limits_{\alpha  = e,\mu ,\tau }  \sum\limits_{i = 1}^3 \bar
\ell_\alpha \gamma^\mu (1 - \gamma^5)U_{\alpha i} \, \nu_i \, W^{-}_\mu
+ \text{h.c.} \right).
\end{equation}
Here $\ell_\alpha$ denotes the field of the charged lepton
$\alpha$, $\nu_i$ denotes the neutrino field with definite mass
$m_i$, and $U_{\alpha i}$ stands for the
Pontecorvo-Maki-Nakagawa-Sakata (PMNS) matrix.

We will also use the Feynman diagram technique in the coordinate
representation, which was developed without reference to S-matrix
theory in paper \cite{Feynman:1949zx}. This approach is more
general than the standard S-matrix approach and is capable of
describing processes passing at finite space and time intervals,
as it will be explained below.

\begin{figure}[htb]
\begin{center}
\includegraphics[width=0.5\linewidth]{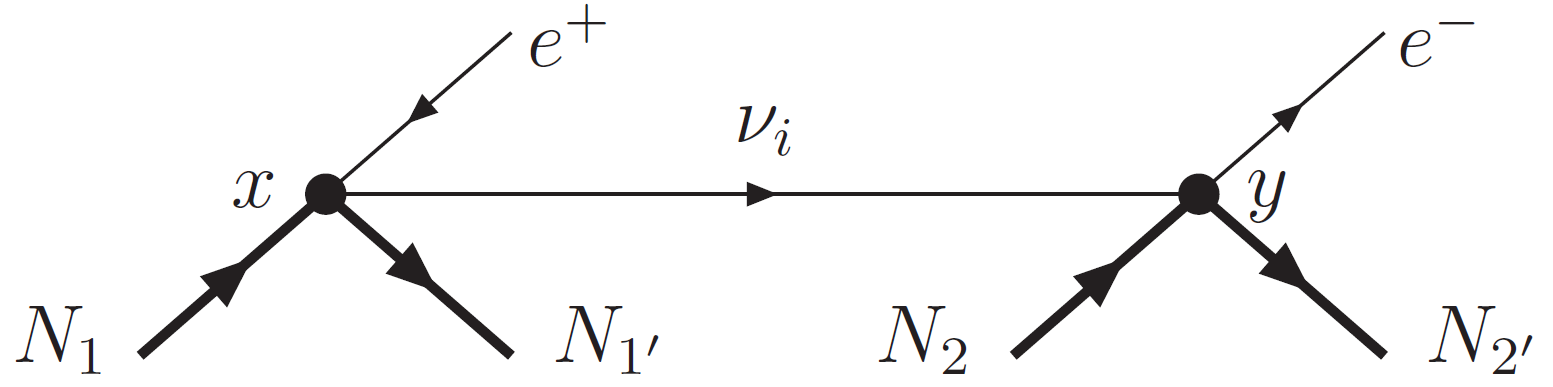}
\caption{Feynman diagram describing the production, propagation, and detection of a massive neutrino as a single process.}
\label{diag1}
\end{center}
\end{figure}

Within this approach, we consider processes, where  neutrinos are
produced and detected in the charged current weak interaction with
nuclei. Namely, we take a process, where a neutrino is produced in
$\beta^+$-decay of nucleus $N_1$ in a source at a point $x$,
propagates over a macroscopic distance to a detector at a point
$y$ and is detected there in the reaction of electron production
at nucleus $N_2$. In the lowest order of perturbation theory with
interaction Lagrangian (\ref{L_cc}) and in the approximation of
Fermi interaction  this process is described by the diagram in Fig.~\ref{diag1} for all the three neutrino mass eigenstates.

We assume that the incoming and outgoing particles
are described by plane waves and have definite momenta. Below it
is shown that all the three virtual neutrino mass eigenstates also
have definite momenta.

The amplitude in the coordinate representation corresponding to the
diagram in Fig.~\ref{diag1} can be easily written out using the Feynman
rules formulated in  textbook \cite{BOSH}. At the moment, we will
not calculate the complete amplitude, we will rather calculate the
part of the diagram with intermediate neutrino $\nu_i$ that
includes only the factors depending on $x$ and $y$ and looks as
follows
\begin{equation}\label{amp_x}
\text{e}^{-\text{i}px} S^{\text{c}}_i(y - x) \text{e}^{\text{i}qy},
\end{equation}
where $S^{\text{c}}_i(y - x)$ is the Feynman propagator of the neutrino
mass eigenstate $\nu_i$, $p$ is the sum of the momenta of the
external lines at $x$ and $q$ is the sum of the momenta of the
external lines at $y$.

According to the prescriptions formulated in paper
\cite{Feynman:1949zx}, to find the amplitude of the process  next
we would have to integrate with respect to $x$ and $y$ over
Minkowski space, which means that we consider the scattering
process to take place in the entire Minkowski space. This would
give us the part of the scattering amplitude corresponding to the
inner fermion line of the diagram in Fig.~\ref{diag1}. However, the
experimental settings suggest that the points $x$ and $y$ are
separated by a directed macroscopic distance $\vec L$. In order to
retain this information, we have to integrate with respect to $x$
and $y$ in such a way that the distance $\vec L$ between these
points remains fixed.  This can be achieved by introducing the
delta function $\delta(\vec y -\vec x - \vec L)$ into the
integral, which enters the amplitude
\begin{equation}\label{amp_L}
\int \text{d}^4x \, \text{d}^4y\, \text{e}^{-\text{i}px} S^{\text{c}}_i(y - x)\, \text{e}^{\text{i}qy}\, \delta(\vec y -\vec x - \vec L).
\end{equation}

Making the change of variables
\begin{equation}\label{chevar}
x = u - \frac{z}{2}, \quad y = u + \frac{z}{2}, \quad
\frac{D(x,y)}{D(u,z)} = 1,
\end{equation}
and integrating  with respect to $u$ in the integral in formula
(\ref{amp_L}), we find that the amplitude is proportional to the
delta function of energy-momentum conservation $(2\pi)^4~\delta(q
- p)$, which will be important for our further considerations. The
remaining integral
\begin{equation}\label{prop_L_mom}
S^{\text{c}}_i(p,\vec L) = \int \text{d}^4z\, \text{e}^{\text{i}pz} S^{\text{c}}_i(z)\,  \delta(\vec z - \vec L)
\end{equation}
will be called the distance-dependent propagator of the neutrino
mass eigenstate $\nu_i$ in the momentum representation.

Obviously, the formal integration of this distance-dependent
fermion  propagator with respect to $\vec L$ over the whole space
gives the standard Feynman fermion propagator in the momentum
representation. Thus, distance-dependent fermion propagator
\eqref{prop_L_mom} can be considered as the spatial density of the
Feynman propagator.

As we have mentioned in the Introduction, this distance-dependent
propagator was first introduced under different assumptions in
paper \cite{Okun1982} and exactly calculated there to be
\begin{equation}\label{prop_L_mom_o}
S^{\text{c}}_i(p,\vec L ) = \frac{\text{e}^{\text{i}(p_i^{\text{ms}} L -\vec p
\vec L )}}{4 \pi L}  \left(\gamma^0 p^0 - \vec \gamma \vec
l\left(p_i^{\text{ms}} + \frac{\text{i}}{L}\right) + m_i\right),
\end{equation}
where $L = | \vec L |$ is the oscillation base, $\vec l = \vec
L/L$  is the unit vector of the direction from a source to a detector,
and $p_i^{\text{ms}} = \sqrt{(p^0)^2 - m_i^2}$.

The distance $L$ between a source and a detector is assumed to be
a macroscopic distance much larger than their sizes. The results
of paper \cite{Grimus:1996av} imply that virtual particles
propagating over macroscopic distances are almost on the mass
shell. This means that $|p^2 - m_i^2|/ \vec p^{\,2} \ll 1$ and we
can expand the square root to the first order in $(p^2 - m_i^2)/
\vec p^{\,2}$. It is also clear that this term can be dropped
everywhere, except in the exponential, where it is multiplied by a
large macroscopic distance $L$. In what follows we will need this
propagator only for $\vec p$ codirectional with $\vec L$. In this case the asymptotic value of the distance-dependent propagator at large macroscopic distances $L$ takes a simple form
\begin{equation} \label{dist_dep_prop_on-shell}
{ S^{\text{c}}_i ( {p, \vec L} ) \simeq  \frac{{\hat p + m_i }}{{4\pi L
}}\,\text{e}^{\text{i}\frac{{p^2 - m_i^2 }}{{2 {|\vec p|}}}L}, }
\end{equation}
which is a kind of spherical wave.

In papers
\cite{Volobuev:2017izt,Egorov:2017qgk,Egorov:2017vdp,Egorov:2019vqv,Egorov:2021iig}
a different definition of distant-dependent propagator was used.
It was defined as
\begin{equation}\label{prop_L_momp}
S^{\text{c}}_i(p,\vec L) = \int \text{d}^4z\, \text{e}^{\text{i}pz} S^{\text{c}}_i(z)\,  \delta(\vec z \vec
l - L),
\end{equation}
and its explicit expression for $\vec p$ codirectional with $\vec L$ was found in paper \cite{Volobuev:2017izt} to be
\begin{equation}\label{prop_L_mom_ap}
S^{\text{c}}_i(p,\vec L) =  \text{i}\,\frac{\hat p + \vec \gamma \vec p\left(1 - \sqrt{1 + \frac{p^2 - m_i^2}{\vec p^{\,2}}}\,\right) + m_i }{2\sqrt{\vec p^{\,2} + p^2 - m_i^2}}\, \text{e}^{-\text{i}\left(|\vec p| - \sqrt{\vec p^{\,2} + p^2 - m_i^2}\,\right) L} \,.
\end{equation}
At macroscopic distances, where the particles are almost on the
mass shell, it also takes a simple form
\begin{equation} \label{dist_dep_prop_on-shell_p}
{ S^{\text{c}}_i ( {p, \vec L}) \simeq  \text{i}\, \frac{{\hat p +
m_i }}{{2 {|\vec p| } }}\,\text{e}^{\text{i}\frac{{p^2 - m_i^2 }}{{2
{|\vec p|}}}L}\, ,}
\end{equation}
which now looks like a kind of plane wave. It is worth noting that
the distant-dependent propagator in this form in the limit
$m_i/|\vec p| \to 0$  was also found in papers
\cite{Fujikawa:2020mei,Fujikawa:2024zol} by a method of path
integration.

Both forms of the distant-dependent propagator can be used for
describing neutrino oscillation processes and lead to the same
results. Here we will use  the form given by formula
(\ref{dist_dep_prop_on-shell}) because it contains the factor
$1/L$, which takes into account the attenuation of the spherical
wave with distance. For the propagator given by formula
(\ref{dist_dep_prop_on-shell_p}), this factor must be added by
hand at the appropriate place. We will comment on this below.

Now we will show how to calculate the probability of a neutrino
oscillation process described by the diagram in Fig.~\ref{diag1} in the
approach under consideration. To this end first we have to assign
4-momenta to the particles in the diagram. We denote the positron
4-momentum by  $q$, the electron 4-momentum by $k$, and the
4-momentum of  the intermediate virtual neutrino by $p$. The
4-momenta of nuclei $N_n$ will be denoted by $P_n$,  $n = 1,
1^\prime, 2, 2^\prime$. The contributions of the nuclei to the
amplitude of the process are completely characterized by the
matrix elements of the weak charged hadron current
\begin{equation}
j_\mu^{(1)} \big( {\vec P_1, \vec P_{1^\prime} } \big) = \big<
N_{1^\prime} \big( { \vec P_{1^\prime} } \big) \big| j_\mu^{(\text{
h})} \big| N_1 \big( {\vec P_1 } \big) \big>, \quad j_\rho^{(2)}
\big( {\vec P_2, \vec P_{2^\prime} } \big) = \big< N_{2^\prime}
\big( { \vec P_{2^\prime} } \big) \big| j_\rho^{(\text{ h})} \big|
N_2 \big( {\vec P_2 } \big) \big>,
\end{equation}
which we will not specify here.

Now we can write out the amplitude of the process using the
standard Feynman rules and replacing the Feynman propagators of
the intermediate neutrinos by the corresponding distance-dependent
propagators (\ref{dist_dep_prop_on-shell}). In the lowest order of
perturbation theory and in the approximation of the Fermi
interaction it reads
\begin{equation}\label{amp_osc}
\begin{split}
M = \, & - \frac{{G_{\text{F}}^2 }}{8\pi L}\sum\limits_{i = 1}^3
{\left| {U_{ei} } \right|^2 \text{e}^{\text{i}\frac{{p^2 - m_i^2 }}{{2 {|\vec p|}}}L} } \\
& \times j_\rho^{(2)} \big( {\vec P_2, \vec P_{2^\prime} } \big)
\, \bar u \big( \vec k \big) \, \gamma ^\rho \left( {1 - \gamma ^5
} \right) (\hat p + m_i) \gamma ^\mu \left( {1 - \gamma ^5 }
\right)v \left( \vec q \right) j_\mu^{(1)} \big( {\vec P_1, \vec
P_{1^\prime} } \big) .
\end{split}
\end{equation}
Here and below we omit the fermion polarization indices for
simplicity.

Since in a $\beta^+$-decay the neutrino energy and momentum are
much larger than the  neutrino masses, we can calculate the
amplitude in the approximation of massless neutrinos. Also taking
into account that the neutrinos are almost on-shell, the squared
modulus of the amplitude, averaged over the polarizations of the
incoming nuclei and summed over the polarizations of the outgoing
particles and nuclei (the operation of averaging and summation is
denoted by the angle brackets), factorizes as follows
\cite{Egorov:2017qgk}
\begin{align}
\label{sqr_amp}
\left\langle {\left| M \right|^2 } \right\rangle  &= \left\langle
{\left| M_{\text{p}} \right|^2 } \right\rangle \left\langle {\left| M_{\text{d}}
\right|^2 } \right\rangle \frac{1}{16\pi^2 L^2} P_{ee} \left( {L}, \left| \vec p \right| \right) ,\\
\label{sqr_M1}
\left\langle {\left| M_{\text{p}} \right|^2 } \right\rangle
&= 4G_{\text{F}}^2 \left( -{g^{\mu \nu } \left( {pq} \right)
+\left( {p^\mu  q^\nu   + q^\mu  p^\nu  } \right) +
\text{i}\varepsilon ^{\mu \nu \alpha \beta } p_{\alpha}  q_\beta  } \right) W_{\mu \nu }^{(1)} , \\
\left\langle {\left| M_{\text{d}} \right|^2 } \right\rangle &= 4G_\text{
F}^2 \left(- {g^{\rho \sigma } \left( {p k} \right) + \left(
{p^\rho  k^\sigma   + k^\rho  p^\sigma  } \right) - \text{i}\varepsilon
^{\rho \sigma \alpha \beta } p_{\alpha} k_\beta  } \right)
W_{\rho \sigma }^{(2)},
\end{align}
where $M_{\text{p}}$ and $M_{\text{d}}$ are the amplitudes of the
neutrino production and detection processes, $W_{\mu \nu }^{(n)} =
\langle {j_\mu ^{(n)} j_\nu^{(n)\dag} } \rangle$, $n=1,2$, are the
nuclear tensors, and we introduced the standard notation
\begin{equation}\label{P_ee}
P_{ee} \left( L, \left| \vec p \right| \right) = 1 -
4\sum\limits_{\scriptstyle i,k = 1  \hfill \atop \scriptstyle i >
k \hfill}^3 {\left| {U_{ei} } \right|^2 \left| {U_{ek} } \right|^2
\sin ^2 \left( {\frac{{ \Delta m_{ik}^2 }}{{4 \left| \vec p
\right|}}L} \right)},\quad \Delta m_{ik}^2 \equiv m_i^2 - m_k^2,
\end{equation}
for the expression, which, in the quantum-mechanical approach, is
called the electron neutrino survival probability.

It is convenient to specify the notations for the energy and
momentum of the intermediate neutrinos, $p = (E, \vec p)$. 
To calculate the probability of the process under consideration we must
multiply squared amplitude (\ref{sqr_amp}) by the delta function
of energy-momentum conservation $(2\pi)^4 \, \delta ( P_1 + P_2 -
P_{1^\prime} - P_{2^\prime} - q - k)$ and to integrate it with
respect to the momenta of the outgoing particles and nuclei. In so
doing we have to keep in mind that the virtual neutrino 4-momentum
$p$ is defined by the energy-momentum conservation in the
production vertex and can be altered by this integration. However,
in neutrino oscillation experiments the admissible values of the
neutrino momentum are restricted by experimental settings, because
the virtual neutrinos propagate from the source to the detector in
the direction of the vector $\vec L$. Therefore, we have to
calculate the differential probability of the process for $\vec p$
codirectional with $\vec L$.

To take this restriction into account, we multiply squared
amplitude (\ref{sqr_amp}) by the integral
\begin{equation} \label{identity}
    \int \delta \left( P_1 - P_{1^\prime} - q - p \right) \text{d}^4p \equiv 1,
\end{equation}
which exists due to the energy-momentum conservation in any vertex
of the diagram in Fig.~\ref{diag1}, as well as by the factor $(2\pi)^3 \,
\delta (\vec p - E \vec l\,)$, which fixes the direction of the
virtual neutrino momentum and is consistent with the used
approximation $p^2 = 0$, and then integrate the obtained
expression with respect to the momenta of the outgoing particles
and nuclei. As a result, we  arrive at the differential
probability of the process, which also factorizes:
\begin{equation} \label{dif_prob_osc}
\begin{split}
\frac{{\text{d} W}}{{\text{d} \Omega_{\vec l}}} = & \, \frac{1}{{2P_1^0 \, 2P_2^0 }}
\int {\frac{{\text{d}^3 k}} {{\left( {2\pi } \right)^3 2k^0 }}\frac{{\text{d}^3 q}}{{\left( {2\pi } \right)^3 2q^0 }}
\frac{{\text{d}^3 P_{1^\prime} }}{{\left( {2\pi } \right)^3 2P_{1^\prime}^0 }}
\frac{{\text{d}^3 P_{2^\prime} }} {{\left( {2\pi } \right)^3 2P_{2^\prime}^0 }} \,
\text{d}^4 p \, {\left\langle {\left| M \right|^2 } \right\rangle } } \\
& \times \left( {2\pi } \right)^4 \delta \left( {P_1 + P_2 - P_{1^\prime} - P_{2^\prime} - q - k} \right)
\delta \left( {P_1 - P_{1^\prime} - q - p} \right) (2\pi)^3 \delta \big( \vec p - E \vec l \ \big) \\
= & \, \frac{{1}}{{L^2}} \int\limits_{E_{\min}}^{E_{\max}} \text{d} E\,
\frac{{\text{d}^3 W_{\text{p}} \left( E \right) }}{{\text{d}E \text{d} \Omega_{\vec
l}}} \,W_{\text{d}} \left( E \right) P_{ee} \left( {E, L} \right).
\end{split}
\end{equation}
Here
\begin{equation}\label{dif_W_1}
\frac{{\text{d}^3 W_{\text{p}} \left( E \right) }}{{\text{d}E \text{d} \Omega_{\vec
l}}} = \frac{1}{{2P_1^0 }} \frac{1}{{\left( {2\pi } \right)^3 2 E
}} \int {\frac{{\text{d}^3 q}} {{\left( {2\pi } \right)^3 2q^0 }}
\frac{{\text{d}^3 P_{1^\prime} }} {{\left({2\pi } \right)^3
2P_{1^\prime}^0 }}  { \left\langle {\left| {M_{\text{p}} }
\right|^2 } \right\rangle } \left( {2\pi } \right)^4 \delta \left(
{P_1 - P_{1^\prime} - q - p} \right)}
\end{equation}
is the differential probability of decay per unit time of  nucleus
$N_1$ into nucleus $N_{1^\prime}$, a positron and a massless
fermion with momentum $\vec p = E \vec l$, which can be viewed as
the electron neutrino of the standard approach,
\begin{equation}\label{dif_W_2}
W_\text{d} \left( E \right)  = \frac{1}{{2P_2^0 \, 2 E }}\int
{\frac{{\text{d}^3 k}}{{\left( {2\pi } \right)^3 2k^0 }} \frac{{\text{d}^3
P_{2^\prime} }}{{\left( {2\pi } \right)^3 2P_{2^\prime}^0}}
{\left\langle {\left| {M_\text{d} } \right|^2 } \right\rangle }
\left( {2\pi } \right)^4 \delta \left( {P_2  + p - P_{2^\prime}  -
k} \right)}
\end{equation}
is the probability of interaction of the massless fermion (the
electron neutrino)  with momentum $\vec p = E \vec l$ and nucleus
$N_2$ with the production of nucleus $N_{2^\prime}$ and an
electron, and  $P_{ee} \left( {E, L} \right)$ is called, in the
quantum-mechanical approach, the electron neutrino survival
probability.

To obtain the same result  for process probability (\ref{dif_prob_osc}) with distance-dependent propagator (\ref{dist_dep_prop_on-shell_p}) one has to fix the direction of the neutrino momentum $\vec p$ by introducing   the factor 
$$ \frac{2\pi}{L^2} \delta (E - |\vec p|) \delta\left (\frac{ \vec p}{E} - \vec l \right) $$
into the integrand, which explicitly takes into account the approximation of massless neutrinos.

If nuclei $N_1$ and $N_2$ are at rest, their energies should be
replaced by their masses. In this case  differential decay
probability (\ref{dif_W_1}) turns out to be  isotropic and can be
written as
\begin{equation}
	\frac{{\text{d}^3 W_{\text{p}} \left( E \right) }}{{\text{d}E \text{d} \Omega_{\vec
	l}}} = \frac{1}{4\pi} \frac{{\text{d} W_{\text{p}} \left( E \right)
	}}{\text{d}E},
\end{equation}
whereas  $W_\text{d}(E)$ coincides with the cross section of the
detection process $\sigma_\text{d}(E)$, because in our system of units
the speed of light $c = 1.$ Finally, the formula
\begin{equation}\label{dif_W_tot}
W_{\text{pd}} = \frac{{1}}{{4\pi L^2}}
\int\limits_{E_{\min}}^{E_{\max}} \text{d} E\, \frac{{\text{d} W_{\text{p}}
\left( E \right) }}{\text{d}E} \, P_{ee} \left( {E,L} \right)
\sigma_{\text{d}} \left( E \right)
\end{equation}
gives the  probability of the process per unit time. The upper
limit of integration is defined by the maximal neutrino energy in
the production process and the lower limit is defined by the
threshold energy of the detection process. To find the average
number of events in the detector per unit time the right-hand side
of this formula should be multiplied by the numbers of nuclei in
the source and detector.

We also introduce the normalized neutrino detection probability
${w_{\text{pd}} }$, which is defined as the ratio
\begin{equation}\label{dif_W_norm}
{w_{\text{pd}} }  = \frac{\int\limits_{E_{\min}}^{E_{\max}} \text{d} E\,
\frac{{\text{d} W_{\text{p}} \left( E \right) }}{\text{d}E} \, P_{ee} \left(
{E,L} \right) \sigma_{\text{d}} \left( E
\right)}{\int\limits_{E_{\min}}^{E_{\max}} \text{d} E\, \frac{{\text{d}
W_{\text{p}} \left( E \right) }}{\text{d}E} \sigma_{\text{d}} \left( E
\right)},
\end{equation}
where the denominator determines the probability of detecting
massless electron neutrinos according to the formula:
\begin{equation}\label{dif_W_el}
W_{\text{pd}}^{(e)} = \frac{{1}}{{4\pi L^2}}
\int\limits_{E_{\min}}^{E_{\max}} \text{d} E\, \frac{{\text{d} W_{\text{p}}
\left( E \right) }}{\text{d}E} \, \sigma_{\text{d}} \left( E \right).
\end{equation}
The normalized probability $w_{\text{pd}}(L)$ characterizes the
behavior of the detection probability with distance and
asymptotically is equal to the relative neutrino deficit in the
process. By fixing the neutrino energy with the help of a delta
functions inserted in the integrands of this formula we get the
normalized differential neutrino detection probability, which, of
course, coincides with $P_{ee} \left( {E,L} \right)$ known as the
electron neutrino survival probability in the standard quantum
mechanical approach. In the next section we will apply these
formulas to particular oscillation processes.

A similar factorization of the process probability also occurs for
proton-proton chain processes with neutrino production in
scattering reactions \cite{Egorov:2021iig}, and consequently,
formula (\ref{dif_W_tot}) essentially works for all solar neutrino
production processes.

It is worth noting here that the approach can also be applied to
describing neutrino oscillation processes, where neutrinos are
detected in interaction with the weak neutral current
\cite{Egorov:2017qgk}, as well as to processes with the production
of other charged leptons in the source \cite{Volobuev:2017rnb}.
Besides neutrino oscillations, the approach also allows one to
describe the oscillations of neutral mesons, which are unstable
particles, as well as the attenuation of the flux of
non-oscillating unstable particles propagating from a source to a
detector \cite{Volobuev:2019zan}.

\section{Coherence length of neutrino oscillations}

Here we will use the derived formulas to describe particular
neutrino oscillation processes, namely the ones, where neutrinos
are produced in the decay
$$
{^{13} \text{N}} \to {^{13} \text{C}} + e^ + + \nu_i,
$$
which is a reaction of the solar carbon cycle, and
detected  by  chlorine-argon or gallium-germanium detectors, i.e.,
in the reactions
$$
\nu_i + {^{37} \text{Cl}} \to {^{37} \text{Ar}} + e^-\quad
\mbox{and}\quad \nu_i + {^{71} \text{Ga}} \to {^{71} \text{Ge}} +
e^-.
$$

In nuclear physics all these processes are referred to the
so-called allowed transitions \cite{Bohr-Mottelson}. This means
that in calculating the amplitudes of these processes one can use
the approximation of free nucleons at rest and neglect the nuclear
form factors. In this case the decay probability of a nucleus
${^{13} \text{N}}$ and the cross sections of the detection
reactions can be written out in the form
\begin{align}
\frac{{\text{d} W_{\text{p}} (E) }}{\text{d} E } \sim & \ \sqrt {\left( {E_{\max }  - E} \right)
\left( {E_{\max }  - E + 2m_e} \right)} \left( {E_{\max }  - E + m_e} \right),  \\
\sigma_{\text{d}} \left( E \right) \sim & \ \sqrt {\left( {E - E_{\min } } \right)
\left( {E - E_{\min }  + 2m_e} \right)} \left( {E - E_{\min } + m_e} \right) .
\end{align}
Here $E_{\max }$ is the maximal neutrino energy determined by the
production process, $E_{\min }$ is the detection threshold energy
determined by the detection process, and $m_e$ is the electron
mass.  For the production and detection processes under
consideration we have:
$$
E_{\max } ({^{13} \text{N}}) = 1199 \text{ keV}, \quad E_{\min } (\text{Ga-Ge})
= 232 \text{ keV}, \quad E_{\min } (\text{Cl-Ar}) = 814 \text{ keV}.
$$

\begin{figure}[htb]
\begin{minipage}[h]{0.49\linewidth}
\begin{center}
\includegraphics[width=1\linewidth]{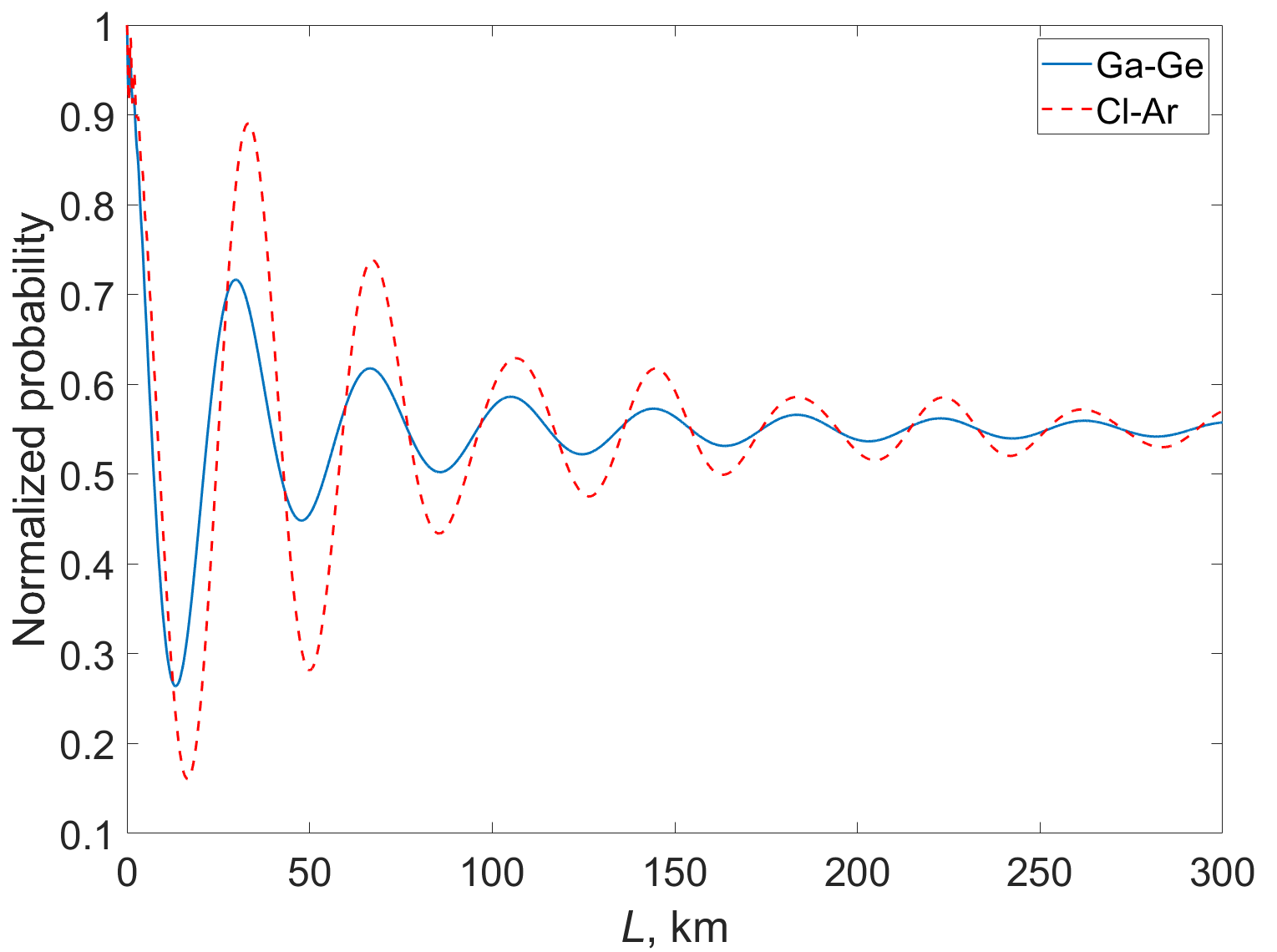} \\ a) Distance $L$ from 0 to 300 km.
\end{center}
\end{minipage}
\hfill
\begin{minipage}[h]{0.49\linewidth}
\begin{center}
\includegraphics[width=1\linewidth]{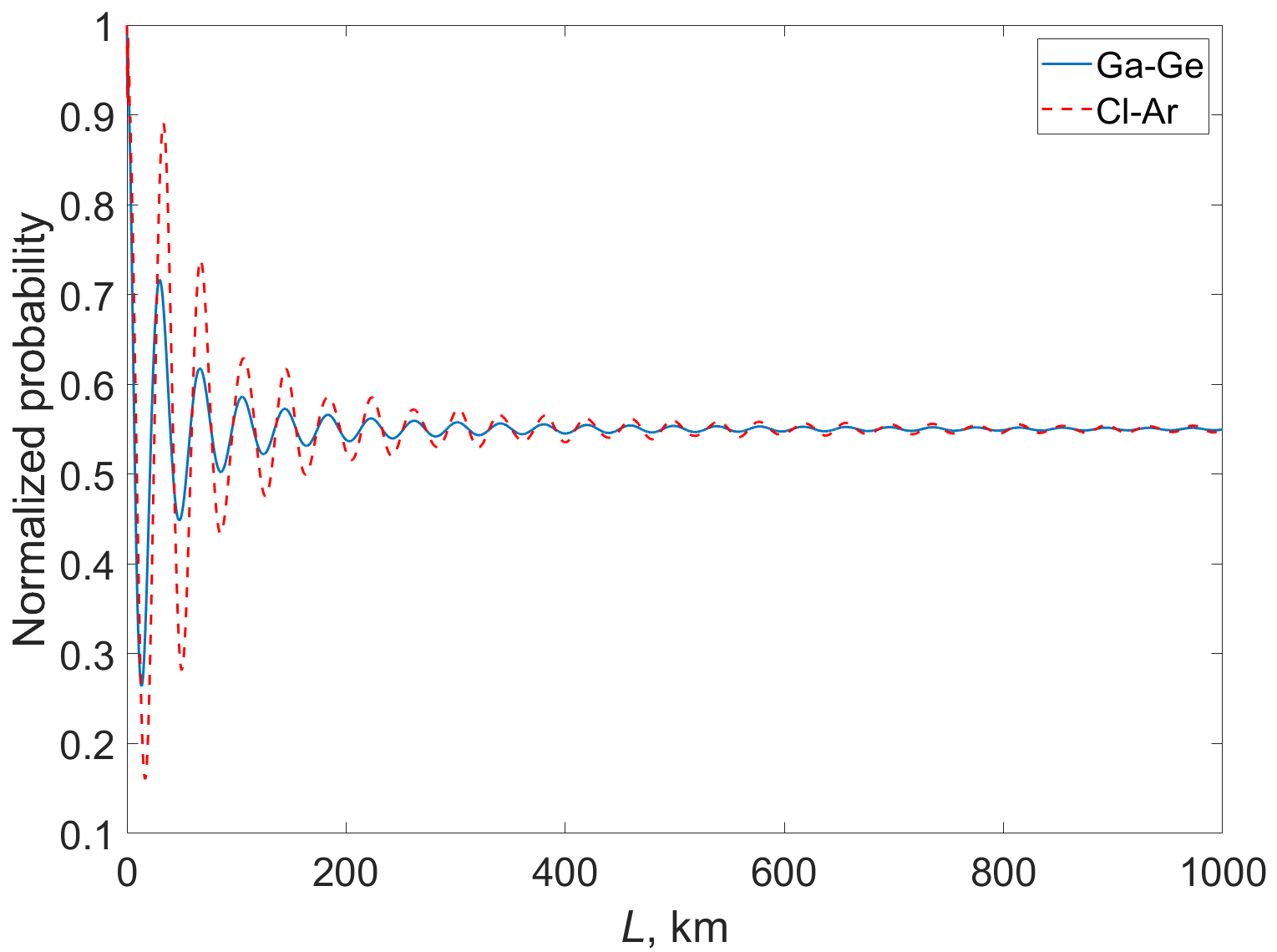} \\ b) Distance $L$ from 0 to 1000 km.
\end{center}
\end{minipage}
\caption{Normalized probabilities of the neutrino oscillation
processes with  the neutrino production in the ${^{13} \text{ N}}$
decay and the registration by Cl-Ar and Ga-Ge detectors.}
\label{fig_osc_1}
\end{figure}

This approximation is rather rough. Nevertheless, it is sufficient
to take into account the spectral characteristics of the
production and detection processes. In accordance with formula
(\ref{dif_W_tot}), the distant-dependent probabilities of the
processes under consideration can be represented in the form
\begin{equation}\label{dif_W_part}
\begin{split}
W_{\text{pd}} =  \frac{{C}}{{4\pi L^2}} \int\limits_{E_{\min}}^{E_{\max}} &
\text{d} E\,\sqrt {\left( {E_{\max } - E} \right)\left( {E_{\max }  - E + 2m_e} \right)}
\left( {E_{\max }  - E + m_e} \right) \\
& \ \times P_{ee} \left( {E,L} \right) \sqrt {\left( {E - E_{\min}
} \right)\left( {E - E_{\min }  + 2m_e} \right)} \left( {E -
E_{\min } + m_e} \right).
\end{split}
\end{equation}
The explicit theoretical expression for the normalization constant
$C$ in terms of $G_\text{F}$ and particle masses, which is  different for
different production and registration processes, is unimportant
for us. Here we are not interested in the whole differential
probability of the processes but rather in the normalized neutrino
detection probability $w_\text{pd}$ (\ref{dif_W_norm}) of these
processes depending on the distance $L$.  It is easy to understand
that this normalized probability can be found by dropping the
factor $1/(4\pi L^2)$ and choosing the value of the constant $C$
in (\ref{dif_W_part}) so that the resulting expression equals 1 at
$L = 0$. The normalized probability calculated numerically in this
manner is presented in Fig.~\ref{fig_osc_1} for two detection
processes. In the calculations, the following values of the
neutrino masses and mixing angles have been used
\cite{Tanabashi:2018oca}:
$$\Delta m_{21}^2 = 7.53 \cdot 10^{ - 5} \text{ eV}^2,
\quad \Delta m_{32}^2 = 2.45 \cdot 10^{ - 3} \text{ eV}^2,$$
$$\sin^2 \theta _{12}  = 0.307, \quad \sin^2\theta _{23}  = 0.545,
\quad \sin^2\theta _{13}  = 2.18 \cdot 10^{-2}.$$

We see that the oscillation pattern depends on the detection
process and  the oscillations fade out with distance, which is due
to the momentum distribution of the intermediate neutrinos and
gives rise to a coherence length in our approach. By analogy with
interference in optics we can introduce the visibility function
\cite{Born}, take the standard condition of oscillations'
visibility and arrive at the coherence lengths
$$L_{\text{coh}}(\text{Ga-Ge})  \approx 76 \text{ km},
\quad L_{\text{coh}}(\text{Cl-Ar})  \approx 158 \text{ km}.$$ In
the Ga-Ge case we have a wider momentum distribution  than in the
Cl-Ar one, hence the Ga-Ge oscillation fade out more rapidly thus
having a smaller coherence length. The largest coherence lengths,
of the order of several tens of thousands of kilometers, are
obtained for the reaction of electron capture by $^7\text{Be}$, since the
resulting neutrinos are nearly monoenergetic
\cite{Egorov:2021iig}.

As one can see in Fig.~\ref{fig_osc_1} the oscillations
asymptotically approach the value close to 0.55. The behavior of
the oscillations at distances much larger than the coherence
length is in fact determined by  oscillations' average with
respect to the distance $L$. Thus, according to
(\ref{dif_W_tot}) and (\ref{P_ee}), the asymptotic behavior of
the oscillations  is given  by the expression
\begin{equation}
P_{ee}^\text{asymp}  = 1 - 4\sum\limits_{\scriptstyle i,k = 1
\hfill \atop \scriptstyle i > k \hfill}^3 {\left| {U_{ei} }
\right|^2 \left| {U_{ek} } \right|^2 \frac{1}{2}}  =
\sum\limits_{i = 1}^3 {\left| {U_{ei} } \right|^4 },
\end{equation}
which approximately equals to 0.55 for the chosen values of the
mixing angles $\theta_{ik}$.

Using this expression for the asymptotic value of $P_{ee}$, we can
rewrite the asymptotic value of process probability
(\ref{dif_W_tot}) as follows:
\begin{equation}\label{dif_W_as}
W_{\text{pd}}^{\text{asymp}} = \frac{{1}}{{4\pi L^2}}
\int\limits_{E_{\min}}^{E_{\max}} \text{d} E\,  \sum\limits_{i = 1}^3
\frac{{\text{d} W_{\text{p},i} \left( E \right) }}{\text{d} E} \,
\sigma_{\text{d},i} \left( E \right) ,
\end{equation}
where
\begin{equation}
\frac{{\text{d} W_{\text{p},i} \left( E \right) }}{\text{d} E} = {\left| {U_{ei}
} \right|^2 } \, \frac{{\text{d} W_{\text{p}} \left( E \right) }}{\text{d} E}
\end{equation}
is the production probability of the $i$-th neutrino mass eigenstate
calculated with Lagrangian (\ref{L_cc}) in the approximation of
Fermi interaction and zero neutrino mass and
\begin{equation}\label{cross_sect_vacuum}
\sigma_{\text{d},i} \left( E \right) = {\left| {U_{ei} } \right|^2
} \, \sigma_{\text{d}} \left( E \right)
\end{equation}
is the scattering cross section of the $i$-th neutrino mass
eigenstate  at a detector nucleus calculated in the same
approximation. This means that at large distances neutrino
oscillations do not influence the probability of the process,
which is just the sum of the products of the detection and
production probabilities of all three neutrino mass eigenstates.
It is interesting to note that this result is compatible with the
reasonings about the behavior of the coherence length in the
standard wave-packet treatment of neutrino oscillations
\cite{Giunti:2007ry}. Thus, we see that, for processes in vacuum,
we immediately get the correct result for the process probability,
if we use the neutrino mass eigenstates instead of the neutrino
flavor states. The question is, whether this assertion remains
valid for solar neutrinos, which are produced and travel a part of their way to Earth in matter.

\section{Oscillations in matter in the approximation of two neutrino mass eigenstates}

\begin{figure}[htb]
\noindent \hfill
\begin{minipage}[h]{0.4\linewidth}
\begin{center}
\includegraphics[width=1\linewidth]{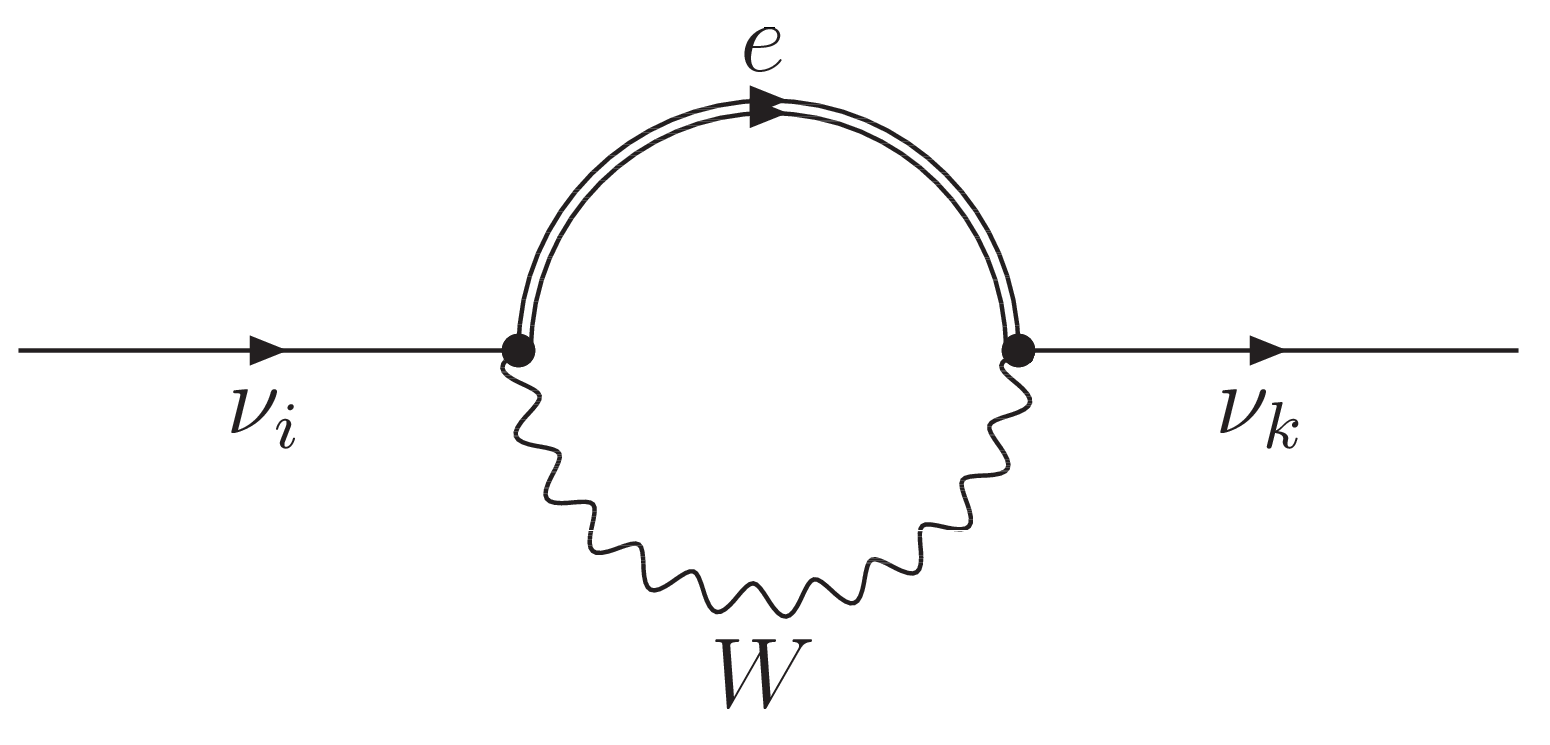} \\
a) Contribution to the self-energy operator due to the interaction
via charged current.
\end{center}
\end{minipage}
\hfill
\begin{minipage}[h]{0.4\linewidth}
\begin{center}
\includegraphics[width=1\linewidth]{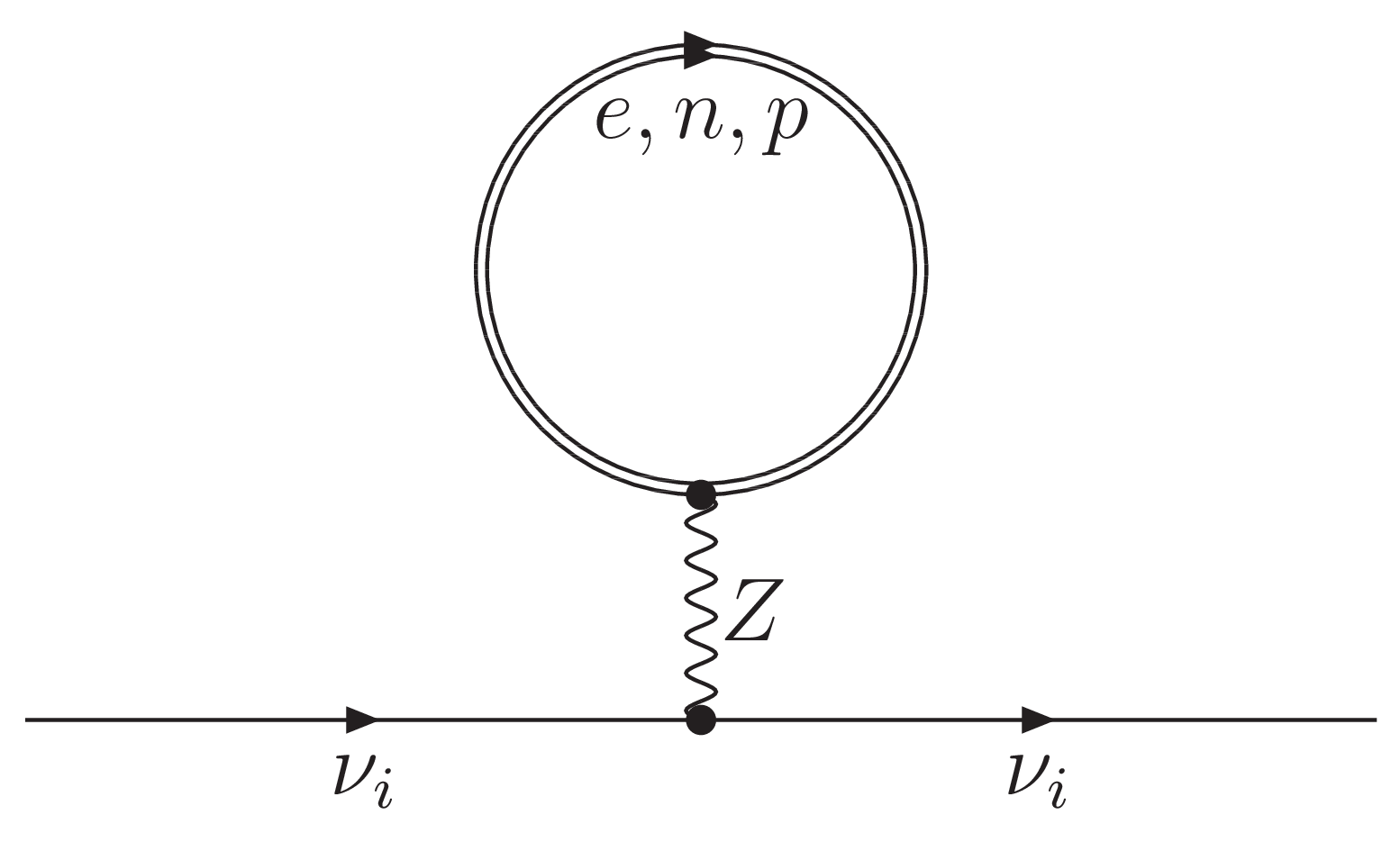} \\
b)  Contribution to the self-energy operator due to the
interaction via neutral current.
\end{center}
\end{minipage}
\hspace{\fill} \caption{Diagrams describing the interactions of
neutrinos with the fermions $e,n,p$ as they propagate through
matter.} \label{matter_diagrams}
\end{figure}

As neutrinos propagate in matter, they interact with the fermions
of the medium through charged and neutral weak currents, which
leads to neutrino mass eigenstate transitions into each other. In
the lowest order of perturbation theory, the process is described
by the self-energy diagrams in Fig.~\ref{matter_diagrams}.

The full Green's function $G(x,y)$ of neutrinos in matter
satisfies the Dyson equation \cite{BOSH} with a neutrino
self-energy operator taking into account the interaction with the
fermions of the medium. For the self-energy operator defined by
the diagrams in Fig. \ref{matter_diagrams} and calculated in the
approximation of the Fermi interaction the Dyson equation takes
the form
\begin{equation}\label{EoM_matter}
\left( {\text{i} I \hat \partial  - M - \frac{1}{2}P_e \hat
f_{\text{cc}}(x) \left( {1 - \gamma ^5 } \right) - \frac{1}{2} I \hat
f_{\text{nc}}(x) \left( {1 - \gamma ^5 } \right)} \right)G(x,y) =
I \delta(x-y),
\end{equation}
where $M = \text{diag} ( m_1 ,m_2, m_3 )$ is the mass matrix, $P_e
= U^\dag \, \text{diag} ( 1,0, 0 ) \, U$ is the projector on the
``electron neutrino state'', $U$ is the PMNS matrix, $I$ denotes
the $3 \times 3$ identity matrix in the space on neutrino states,
$f_\text{cc}^\mu (x)$ and $f_\text{nc}^\mu (x)$ are the effective
charged- and neutral-current potentials of interaction with
matter, which are expressed through the Fermi constant and the
4-vectors of current and polarization of the fermions in the
medium \cite{Lobanov:2016qug}. Green's function $G(x,y)$, which
describes the propagation of all tree types of neutrinos as a
whole, is a $12 \times 12$ matrix.

If the medium is homogeneous and non-polarized, the effective
potentials are constant and proportional to each other
\cite{Chukhnova:2018ddf}:
\begin{equation}
    f_{\text{cc}}^\mu   = f^\mu  , \qquad f_{\text{nc}}^\mu   =
    af^\mu.
\end{equation}
The translation invariance in the medium is restored and  Green's
function $G(x,y) = G(y-x).$ Now it is convenient to transform
equation (\ref{EoM_matter}) to the momentum representation
\begin{equation} \label{eq_GF}
\left( \left(\hat p - \frac{a}{2}\hat f\left( {1 - \gamma ^5 } \right)\right) I - \frac{1}{2}\hat f\left( {1 - \gamma ^5 } \right) P_e - M \right) G(p) = I.
\end{equation}
Thus, we consider Green's function $G(p)$ and all the other matrices
in the equation as tensor products of $3 \times 3$ matrices and
$4 \times4 $ matrices taking values in the Dirac algebra, i.e.\ as
$3 \times3 $ matrices with entries from the Dirac algebra. With this in mind, we will not write $I$ explicitly in what follows.

We will look for $G(p)$ in the form
\begin{equation} \label{subst_GF}
    G(p) = \left( \hat p - \frac{a}{2}\hat f\left( {1 - \gamma ^5 } \right) - \frac{1}{2}\hat f\left( {1 - \gamma ^5 } \right) P_e + M \right) \tilde G(p).
\end{equation}
The equation for $\tilde G(p)$ looks as follows:
\begin{equation} \label{eq_TGF}
\left[\left( \hat p - \frac{1}{2}\hat f\left( {1 - \gamma ^5 } \right) \left( a + P_e \right) \right)^2 + \frac{1}{2}\hat f\left( {1 - \gamma ^5 } \right) \left[P_e,M \right]  -  M^2 \right] \tilde G(p) =  1.
\end{equation}
The term $\hat f\left( {1 - \gamma ^5 } \right) \left[P_e,M
\right]/2$ is of the order $f^\mu$ multiplied by a neutrino mass.
We will assume that $f^\mu \ll m_i$, which is valid for the solar
matter. Thus, the term is very small and can be dropped. In this
approximation, the equation for $\tilde G(p)$ can be rewritten as
\begin{equation} \label{eq_TGF1}
\left[p^2 - pf \left(1 - S\right) \left( a + P_e \right)  -  M^2 \right] \tilde G(p) =  1,
\end{equation}
where
\begin{equation}
S = \frac{1}{{2\sqrt {(pf)^2  - p^2 f^2 } }}\big[ {\hat p,\hat f}
\big]\gamma ^5
\end{equation}
is a generalized helicity operator \cite{Chukhnova:2018ddf}, whose
eigenvalues are equal to $\pm 1$, so that the operators $(1 \pm
S)/2$ are projectors. Representing $\tilde G(p)$ in the form
\begin{equation}
\tilde G(p) = \tilde G_+(p) + \tilde G_-(p), \qquad \tilde G_\pm(p)
= \frac{1 \pm S}{2} \tilde G(p),
\end{equation}
we can solve equation \eqref{eq_TGF1} in the subspaces of the eigenvalues of
$S$ independently and then find a solution for Green's function
$G(p)$ from representation (\ref{subst_GF}).

Here we will do this in the approximation of two neutrino mass eigenstates, since this case is technically easier and more physically transparent, which is good for a demonstration. This significantly helps us better understand the physics of neutrino production and propagation processes in matter.

In this approximation we have to put  $M = \text{diag} ( m_1 ,m_2
)$ and $P_e = U^\dag \, \text{diag} ( 1,0 ) \, U$ in equation
\eqref{EoM_matter}. We also denote $\Delta m_{21}^2 \equiv \Delta
m^2$. As we have explained, a good approximation for  Green's
function of equation \eqref{EoM_matter} in the momentum
representation is
\begin{equation} \label{Green}
    G\left( p \right) = \left( {\hat p - \frac{1}{2}\hat f\left( {1 - \gamma ^5 } \right)
    \left( {a + P_e } \right) + M} \right)\left( \tilde G_+\left( p \right) + \tilde G_- \left( p \right)
    \right).
\end{equation}
The functions $G_\pm\left( p \right)$ can be easily found from
equation (\ref{eq_TGF1}) and turn out to be
\begin{equation}
    \tilde G_ +  \left( p \right) = \text{diag} \left( {\frac{1}{{p^2  - m_1^2 }},\frac{1}{{p^2  - m_2^2 }}} \right)
\end{equation}
and
\begin{equation} \label{left_prop}
\tilde G_ -  \left( p \right) = V \, \text{diag} \left( {\frac{1}{{p^2
- 2a\,pf - m_{1\text{M}}^2 }},\frac{1}{{p^2  - 2a\,pf -
m_{2\text{M}}^2 }}} \right) V^\dag,
\end{equation}
where the unitary matrix
\begin{equation} \label{V}
    V = \left( {\begin{array}{*{20}c}
   {\cos \left( {\theta _\text{M}  - \theta } \right)} & {\sin \left( {\theta _\text{M}  - \theta } \right)}  \\
   { - \sin \left( {\theta _\text{M}  - \theta } \right)} & {\cos \left( {\theta _\text{M}  - \theta } \right)}  \\
\end{array}} \right)
\end{equation}
performs the rotation of the neutrino basis from the mass
eigenstates  in vacuum to the mass eigenstates in the medium and
$m_{i\text{M}}^2$ are the effective masses squared of neutrinos in
matter defined by the eigenvalues of the Hermitian matrix
$2pf\,P_e  + M^2$. This matrix is diagonalized  by the matrix $V$:
\begin{equation}
    V^\dag  \left( {2pf\,P_e  + M^2 } \right)V = \text{diag}\left( {m_{1\text{M}}^2 ,m_{2\text{M}}^2 } \right),
\end{equation}
\begin{equation}
    m_{1\text{M}}^2  = m_1^2  + pf + \frac{{\Delta m^2 }}{2} - \frac{{\Delta m_\text{M}^2 }}{2}, \quad m_{2\text{M}}^2  = m_2^2  + pf - \frac{{\Delta m^2 }}{2} + \frac{{\Delta m_\text{M}^2 }}{2},
\end{equation}
\begin{equation}
    \Delta m_\text{M}^2  = m_{2\text{M}}^2  - m_{1\text{M}}^2  = \sqrt {\left( {\Delta m^2 \cos 2\theta  - 2pf} \right)^2  + \left( {\Delta m^2 } \right)^2 \sin ^2 2\theta } .
\end{equation}
The effective mixing angle in matter $\theta _\text{M}$  entering
formula \eqref{V} is given by
\begin{equation} \label{mixing_angle}
\cos 2\theta _\text{M}  = \frac{{\Delta m^2 \cos 2\theta  -
2pf}}{{\Delta m_\text{M}^2 }}, \quad \sin 2\theta _\text{M}  =
\frac{{\Delta m^2 \sin 2\theta }}{{\Delta m_\text{M}^2 }}.
\end{equation}
These expressions for the effective masses and angle  in matter
coincide with the standard ones \cite{Giunti:2007ry}.

Now let us consider the case of  matter at rest, where
\begin{equation} \label{matter_at_rest}
    f^\mu   = (V_\text{cc} ,\vec 0) \quad \Rightarrow \quad 2pf = 2EV_\text{cc}  \equiv A_\text{cc}
    , \quad V_\text{cc} = \sqrt{2} \, G_\text{F} \, n_{(e)},
\end{equation}
$n_{(e)}$ denoting the electron density.

Next, we substitute Green's function \eqref{Green} into the
definition of distance-dependent propagator \eqref{prop_L_mom}.
Assuming that neutrinos are ultrarelativistic  with $\vec p$
codirectional with $\vec L$ and almost on-shell we obtain the
distance-dependent neutrino propagator in matter in the form
\begin{equation} \label{ddp_matter}
    G(p, L) = G_ +  (p, L) + G_ -  (p, L),
\end{equation}
where
\begin{equation}
G_ +  (p,L) = \frac{{1 + \gamma ^5 }}{2}\frac{{\hat p}}{{4\pi L}} \,
\text{diag} \left( {\text{e}^{\text{i}\frac{{p^2  - m_1^2 }}{{2\left|
{\vec p} \right|}}L} ,\text{e}^{\text{i}\frac{{p^2  - m_2^2
}}{{2\left| {\vec p} \right|}}L} } \right)
\end{equation}
gives the contribution of the sterile right neutrinos and
\begin{equation} \label{ddp_minus}
G_ -  (p,L) = \frac{{1 - \gamma ^5 }}{2} V G_ - ^\text{M}  (p,L)
V^\dag  
\end{equation}
gives the contribution of the active left neutrinos, the
distance-dependent propagator of the neutrino mass eigenstates in
matter having the form
\begin{equation} \label{ddp_M}
G_ - ^\text{M}  (p,L) = \frac{{\hat p}}{{4\pi L}} \, \text{diag}
\left( {\text{e}^{\text{i}\frac{{p^2  - aA_{\text{cc}}  -
m_{1\text{M}}^2 }}{{2\left| {\vec p} \right|}}L} ,\text{e}^{\text{
i}\frac{{p^2  - aA_{\text{cc}}  - m_{2\text{M}}^2 }}{{2\left|
{\vec p} \right|}}L} } \right).
\end{equation}
In what follows, we will drop the term $G_ +  (p,L)$, because  it
does not contribute to process amplitudes, as one will see
shortly.

Let us consider a neutrino oscillation process in matter described
by a diagram similar to the one in Fig.~\ref{diag1}. Now we do not
require the matter density to be constant along the entire
neutrino path but assume it to change slowly enough so that the
adiabaticity condition is satisfied \cite{Giunti:2007ry}:
\begin{equation}
2E\sin 2\theta _\text{M} \left| {\frac{{\text{d}A_{\text{cc}}
}}{{\text{d}z}}} \right| \ll \left( {\Delta m_\text{M}^2 }
\right)^2
\end{equation}
(here $z$ is the coordinate along the source-detector line). This
condition means that there are no transitions from one neutrino
mass eigenstate in matter into another, so that these mass
eigenstates evolve independently.

Since it is the neutrino mass eigenstates in matter that propagate
through the medium in the QFT approach under consideration, in the
adiabatic case we can use propagator \eqref{ddp_M} of the neutrino
eigenstates in matter (or, equivalently, full propagator
\eqref{ddp_matter} in the matter basis, $V^\dag G(p, L) V$),
substituting there the averaged characteristics of the medium.
However, the matrices $V$ in the interaction vertices, which
depend on the characteristics of matter, must be taken local,
i.e.\ for the values of the parameters at the corresponding
points. We will denote these matrices by $V^{(\text{i})}$ and
$V^{(\text{f})}$ for the initial and final points of the neutrino
path. Thus, the distance-dependent propagator of neutrinos in
matter in the adiabatic approximation can be written as
\begin{equation} \label{ddp_adiab}
G^{\text{adiab}} \left( {p,L} \right) = \frac{{1 + \gamma ^5 }}{2}
\cdot \left[ {...} \right] + \frac{{1 - \gamma ^5 }}{2} \cdot
V^{(\text{f})} \bar G_ - ^\text{M} \left( {p,L}
\right)V^{(\text{i})\dag }
\end{equation}
or, explicitly,
\begin{equation}
\begin{split}
    & G^{\text{adiab}} \left( {p,L} \right) = \frac{{1 + \gamma ^5 }}{2}\left[ {...} \right] \\
    & \quad
    + \frac{{1 - \gamma ^5 }}{2}\frac{{\hat p}}{{4\pi L}}\left[ {\left( {\begin{array}{*{20}c}
   {c_\text{i} c_\text{f} } & { - s_\text{i} c_\text{f} }  \\
   { - c_\text{i} s_\text{f} } & {s_\text{i} s_\text{f} }  \\
\end{array}} \right)\text{e}^{\text{i}\frac{{p^2  - \overline {aA_{\text{cc}} }  - \overline {m_{1\text{M}}^2 } }}{{2\left| {\vec p} \right|}}L}  + \left( {\begin{array}{*{20}c}
   {s_\text{i} s_\text{f} } & {c_\text{i} s_\text{f} }  \\
   {s_\text{i} c_\text{f} } & {c_\text{i} c_\text{f} }  \\
\end{array}} \right)\text{e}^{\text{i}\frac{{p^2  - \overline {aA_{\text{cc}} }  - \overline {m_{2\text{M}}^2 } }}{{2\left| {\vec p} \right|}}L} } \right],
\end{split}
\end{equation}
where the ellipsis denotes the irrelevant contribution of the
right neutrinos, $c_n  = \cos \big( {\theta _\text{M}^{(n)}  -
\theta } \big)$, $s_n  = \sin \big( {\theta _\text{M}^{(n)}  -
\theta } \big)$, $n = \text{i}, \text{f}$; $\theta
_\text{M}^{(n)}$ being the effective mixing angle
\eqref{mixing_angle} at the production or detection point, and the
line above an expression denotes taking the corresponding
expression averaged over the neutrino path.

The amplitude of the process in the momentum representation reads
\begin{equation} \label{amplitude}
\begin{split}
    M = & \,  - \frac{{G_\text{F}^2 }}{2}j_\rho ^{(2)} \big(\vec P_2 ,\vec P_{2'} \big)\,\bar u\big(\vec k\big)\,\gamma ^\rho  \left( {1 - \gamma ^5 } \right) \\
    & \,  \times \sum\limits_{i,k = 1}^3 {\left[ {U_{ei} G_{ik}^{\text{adiab}} \left( {p,L} \right)U_{ek}^ *  } \right]} \,\gamma ^\mu  \left( {1 - \gamma ^5 } \right)v\left(\vec q\right)\,j_\mu ^{(1)} \big(\vec P_1 ,\vec P_{1'} \big) \\
  = & \,  - \frac{{G_\text{F}^2 }}{{8\pi L}}j_\rho ^{(2)} \big(\vec P_2 ,\vec P_{2'} \big)\,\bar u\big(\vec k\big)\,\gamma ^\rho  \left( {1 - \gamma ^5 } \right)\hat p\,\gamma ^\mu  \left( {1 - \gamma ^5 } \right)v\left(\vec q\right)\,j_\mu ^{(1)} \big(\vec P_1 ,\vec P_{1'} \big)  \\
  & \, \times \left( {\cos \theta _\text{M}^{(\text{i})} \cos \theta _\text{M}^{(\text{f})} \text{e}^{\text{i}\frac{{p^2  - \overline {aA_{\text{cc}} }  - \overline {m_{1\text{M}}^2 } }}{{2\left| {\vec p} \right|}}L}  + \sin \theta _\text{M}^{(\text{i})} \sin \theta _\text{M}^{(\text{f})} \text{e}^{\text{i}\frac{{p^2  - \overline {aA_{\text{cc}} }  - \overline {m_{2\text{M}}^2 } }}{{2\left| {\vec p} \right|}}L} } \right).
\end{split}
\end{equation}
It is obvious that the term proportional to $1 + \gamma ^5$ in
\eqref{ddp_adiab}   does not contribute to the amplitude.

Carrying out the procedure described in Section 2, we arrive at
the expression for the probability per unit time of the process
under consideration:
\begin{equation} \label{Wpd_matter}
    W_{\text{pd}}  = \frac{1}{{4\pi L^2 }}\int\limits_{E_{\min } }^{E_{\max } } {\text{d}E \, \frac{{\text{d}W_\text{p} \left( E \right)}}{{\text{d}E}} \, P_{ee}^{\text{M, adiab}} \left( {E,L} \right)  \sigma _\text{d} \left( E \right) },
\end{equation}
where
\begin{equation} \label{PeeM}
    P_{ee}^{\text{M, adiab}} \left( {E,L} \right) = \frac{1}{2} + \frac{1}{2}\cos 2\theta _\text{M}^{(\text{i})} \cos 2\theta _\text{M}^{(\text{f})}  + \frac{1}{2}\sin 2\theta _\text{M}^{(\text{i})} \sin 2\theta _\text{M}^{(\text{f})} \cos \left( {\frac{{\Delta \overline {m_\text{M}^2 } }}{{2E}}L} \right)
\end{equation}
is the standard expression for the oscillations of two types  of
neutrinos in matter in the adiabatic approximation
\cite{Giunti:2007ry}, and $\Delta \overline {m_\text{M}^2 } =
\overline {m_{2\text{M}}^2 } - \overline {m_{1\text{M}}^2 }$. For
large distances $L$, where the oscillations are averaged out due
to the spread of the neutrino energy or the extension of the
source, we obtain the standard asymptotic formula
\begin{equation} \label{PeeM_asymp}
P_{ee}^{\text{M, adiab, asymp}} \left( E \right) = \frac{1}{2} + \frac{1}{2}\cos
2\theta _\text{M}^{(\text{i})} \cos 2\theta _\text{M}^{(\text{f})}
= \cos^2\theta _\text{M}^{(\text{i})} \cos^2\theta
_\text{M}^{(\text{f})} +  \sin^2 \theta _\text{M}^{(\text{i})}
\sin^2 \theta _\text{M}^{(\text{f})}.
\end{equation}
This means that, in the approximation of two neutrino mass
eigenstates, we can write the asymptotic neutrino detection
probability in matter in the same form as in vacuum:
\begin{equation}\label{dif_W_asM}
W_{\text{pd}}^{\text{asymp}} = \frac{{1}}{{4\pi L^2}}
\int\limits_{E_{\min}}^{E_{\max}} \text{d} E\,  \sum\limits_{i = 1}^2
\frac{{\text{d} W_{\text{p},i} \left( E \right) }}{\text{d} E} \,
\sigma_{\text{d},i} \left( E \right) ,
\end{equation}
where
\begin{equation}
\frac{{\text{d} W_{\text{p},1} \left( E \right) }}{\text{d} E} = \cos^2\theta
_\text{M}^{(\text{i})} \, \frac{{\text{d} W_{\text{p}} \left( E \right)
}}{\text{d} E}, \qquad \frac{{\text{d} W_{\text{p},2} \left( E \right) }}{\text{d} E} =
\sin^2\theta _\text{M}^{(\text{i})} \, \frac{{\text{d} W_{\text{p}}
\left( E \right) }}{\text{d} E}
\end{equation}
are the production probabilities of the neutrino eigenstates in
the matter at $L =0$ in the approximation of the Fermi interaction
and zero neutrino mass and
\begin{equation}
\sigma_{\text{d},1} \left( E \right) = \cos^2\theta
_\text{M}^{(\text{f})} \, \sigma_{\text{d}} \left( E \right),\quad
\sigma_{\text{d},2} \left( E \right) = \sin^2\theta
_\text{M}^{(\text{f})} \, \sigma_{\text{d}} \left( E \right)
\end{equation}
are the scattering cross sections of the neutrino eigenstates  in
the matter at $L$ on a detector nucleus  calculated in the same
approximation.

Thus, we see that neutrino oscillations in matter do not play any
role in describing processes, where neutrinos are detected at
distances much larger than the coherence length. In the next
section we will show that it is exactly this scenario that is
realized in experiments with solar neutrinos.

\section{Solar neutrinos}

Now that we have demonstrated, how our approach works in the physically transparent and more convenient for analytical treatment two-neutrino case, we move on to describing solar neutrino processes taking into account all three neutrino mass eigenstates, which is a more involved task.

First of all, we again need to diagonalize the matrix $2pf\,P_e  +
M^2$. The equation for the effective masses squared in matter,
which are eigenvalues of this matrix, has the form
\begin{equation} \label{mM2_eq}
    \det \left( {2pf\,P_e  + M^2  - m_{i\text{M}}^2 } \right) = 0.
\end{equation}
This equation is cubic in $m_{i\text{M}}^2$ and can be solved
using  Cardano's formula. The solution is cumbersome and poorly
suited for analytical study but it can be considered known and
used in numerical modelling. The explicit form of equation
\eqref{mM2_eq} and its analytical solutions can be found in
Appendix. The diagonalizing matrix $V$ is composed of the
normalized eigenvectors of $2pf\,P_e  + M^2$ and has the form
\begin{equation} \label{V_3f}
V = \left( {\begin{array}{*{20}c}
{\frac{1}{{N_1 }}\delta m_{21}^2 \delta m_{31}^2 U_{e1}^ * } & { - \frac{1}{{N_2 }} \delta m_{22}^2 \delta m_{32}^2 U_{e1}^ * } & {\frac{1}{{N_3 }} \delta m_{23}^2 \delta m_{33}^2 U_{e1}^ * }  \\
{\frac{1}{{N_1 }} \delta m_{11}^2 \delta m_{31}^2 U_{e2}^ * } & { - \frac{1}{{N_2 }} \delta m_{12}^2 \delta m_{32}^2 U_{e2}^ *} & {\frac{1}{{N_3 }} \delta m_{13}^2 \delta m_{33}^2 U_{e2}^ * }  \\
{\frac{1}{{N_1 }} \delta m_{11}^2 \delta m_{21}^2 U_{e3}^ * } & { - \frac{1}{{N_2 }} \delta m_{12}^2 \delta m_{22}^2 U_{e3}^ * } & {\frac{1}{{N_3 }} \delta m_{13}^2 \delta m_{23}^2 U_{e3}^ * }  \\
\end{array}} \right),
\end{equation}
where $\delta m_{ki}^2  = m_k^2  - m_{i\text{M}}^2$ , and the
normalization constant of the $i$-th eigenvector is
\begin{equation} \label{Ni}
N_i  = \sqrt {\left( {\delta m_{2i}^2 } \right)^2 \left( {\delta
m_{3i}^2 } \right)^2 \left| {U_{e1} } \right|^2  + \left( {\delta
m_{1i}^2 } \right)^2 \left( {\delta m_{3i}^2 } \right)^2 \left|
{U_{e2} } \right|^2  + \left( {\delta m_{1i}^2 } \right)^2 \left(
{\delta m_{2i}^2 } \right)^2 \left| {U_{e3} } \right|^2 } .
\end{equation}

The Green's function of equation \eqref{EoM_matter} for neutrinos
in matter has the same form \eqref{Green} in the case of three
neutrino mass eigenstates, and it is obvious that the right
neutrino contribution to it reads
\begin{equation}
    \tilde G_ +  \left( p \right) = \text{diag} \left( {\frac{1}{{p^2  - m_1^2 }},\frac{1}{{p^2  - m_2^2 }},\frac{1}{{p^2  - m_3^2 }}} \right),
\end{equation}
while the left neutrino contribution is
\begin{equation}
    \tilde G_ -  \left( p \right) = V \, \text{diag} \left( {\frac{1}{{p^2  - 2a\,pf - m_{1\text{M}}^2 }},\frac{1}{{p^2  - 2a\,pf - m_{2\text{M}}^2 }},\frac{1}{{p^2  - 2a\,pf - m_{3\text{M}}^2 }}} \right) V^\dag .
\end{equation}

Having the Green's function in matter, one can calculate the
distance-dependent propagator \eqref{prop_L_mom}. Again, assuming
that the matter is at rest \eqref{matter_at_rest} and the
neutrinos are ultrarelativistic, almost on-shell, and $\vec p
\uparrow\uparrow \vec L$, one arrives at the expression of the
form \eqref{ddp_matter}, \eqref{ddp_minus}, where
\begin{align}
& G_ +  (p,L) = \frac{{1 + \gamma ^5 }}{2}\frac{{\hat p}}{{4\pi L}} \, \text{diag}
\left( {\text{e}^{\text{i}\frac{{p^2  - m_1^2 }}{{2\left| {\vec p} \right|}}L}
,\text{e}^{\text{i}\frac{{p^2  - m_2^2 }}{{2\left| {\vec p} \right|}}L},
\text{e}^{\text{i}\frac{{p^2  - m_3^2 }}{{2\left| {\vec p} \right|}}L} } \right), \\
\label{ddp_M_3f} & G_ - ^\text{M}  (p,L) = \frac{{\hat p}}{{4\pi
L}} \, \text{diag} \left( {\text{e}^{\text{i}\frac{{p^2  -
aA_{\text{cc}}  - m_{1\text{M}}^2 }}{{2\left| {\vec p} \right|}}L}
,\text{e}^{\text{i}\frac{{p^2  - aA_{\text{cc}}  - m_{2\text{M}}^2
}}{{2\left| {\vec p} \right|}}L} , \text{e}^{\text{i}\frac{{p^2  -
aA_{\text{cc}}  - m_{3\text{M}}^2 }}{{2\left| {\vec p} \right|}}L}
} \right).
\end{align}

In the case of three neutrino mass eigenstates, the adiabaticity
condition is written as
\begin{equation}
2E \left| {\frac{{\text{d}A_{\text{cc}} }}{{\text{d}z}}} \right| \ll
\left( {\Delta m_{ik\text{M}}^2 } \right)^2 ,
\end{equation}
and it is known to be valid for the density of the solar matter
\cite{Giunti:2007ry}. Then the distance-dependent propagator in
the adiabatic approximation is given by expression
\eqref{ddp_adiab}, where $\bar G_ - ^\text{M} \left( {p,L}
\right)$ denotes the propagator \eqref{ddp_M_3f} of the mass
eigenstates in matter, taken with the averaged characteristics of
the medium, and $V^{(\text{i})}$ and $V^{(\text{f})}$ are the
diagonalizing matrices \eqref{V_3f}, calculated with the
parameters of the medium at the beginning and at the end of the
neutrino path, respectively.

Writing down the amplitude of the process in matter similar to
that described by the diagram in Fig.~\ref{diag1}, we obtain an
expression similar to \eqref{amplitude}, where the oscillating
factor can be written in the form
\begin{equation}
\sum\limits_{i,k,j = 1}^3  {U_{ei} V_{ij}^{(\text{f})} \, \text{e}^{\text{i}\,\frac{{p^2  - \overline {aA_{\text{cc}} }  -
\overline {m_{j\text{M}}^2 } }}{{2\left| {\vec p} \right|}}\,L} \,
V_{kj}^{(\text{i}) * } U_{ek}^ *  }   = \sum\limits_{i = 1}^3
{K_{ei}^{(\text{i})} K_{ei}^{(\text{f})} \,\text{e}^{\text{i}\,\frac{{p^2
- \overline {aA_{\text{cc}} }  - \overline {m_{i\text{M}}^2 }
}}{{2\left| {\vec p} \right|}}\,L} }
\end{equation}
with
\begin{equation} \label{Ki}
K_{ei}^{(n)} = \sum\limits_{k = 1}^3 { U_{ek} V_{ki} }  = \left.
{\frac{{\left( { - 1} \right)^{i + 1} }}{{N_i }}\left( {\left|
{U_{e1} } \right|^2 \delta m_{2i}^2 \delta m_{3i}^2  + \left|
{U_{e2} } \right|^2 \delta m_{1i}^2 \delta m_{3i}^2  + \left|
{U_{e3} } \right|^2 \delta m_{1i}^2 \delta m_{2i}^2 } \right)}
\right|^{(n)} ,
\end{equation}
$n = \text{i},\text{f}$. Finally, one arrives at an expression for
the process probability per unit time in the form
\eqref{Wpd_matter}, where the distance-dependent  normalized
differential neutrino detection probability (which is called the
electron neutrino survival probability in the standard approach)
reads
\begin{equation} \label{PeeM_adiab}
P_{ee}^{\text{M, adiab}} \left( {E,L} \right) = \sum\limits_{i =
1}^3 {\big( {K_{ei}^{(\text{i})} } \big)^2 \big(
{K_{ei}^{(\text{f})} } \big)^2 }  + 2\sum\limits^3_{\scriptstyle
i,k = 1 \hfill \atop \scriptstyle i > k \hfill}
{K_{ei}^{(\text{i})} K_{ei}^{(\text{f})} K_{ek}^{(\text{i})}
K_{ek}^{(\text{f})} \cos \left( {\frac{{\Delta \overline
{m_{ik\text{M}}^2 } }}{{2E}}\,L} \right)} .
\end{equation}
The asymptotic value of the normalized  probability for large $L$
is given by
\begin{equation} \label{PeeM_adiab_asymp}
P_{ee}^{\text{M, adiab, asymp}} \left( {E} \right) =
\sum\limits_{i = 1}^3 {\big( {K_{ei}^{(\text{i})} } \big)^2 \big(
{K_{ei}^{(\text{f})} } \big)^2 }.
\end{equation}
From definitions \eqref{Ki} and \eqref{Ni} of the quantities
$K_{ei}^{(n)}$ and $N_i$ it is easy to see that $K_{ei}^{(n)} \to
\left| {U_{ei} } \right|$, when $A_\text{cc} \to 0$.

Typically, the results of solar neutrino experiments are presented
in a compact form in terms of the experimentally observed electron
neutrino survival probability $P_{ee}(E)$ (called the normalized differential neutrino detection probability in our approach), which is defined as the ratio of
the number of neutrinos detected in experiments with neutrino
registration through charged current weak interaction  in a small
energy range to the number of electron neutrinos that would  be
detected  in that energy range according to the standard solar
model without neutrino oscillations \cite {Borexino}. It is due to
this definition that we consider only the detection processes with
the charged current weak interaction, since the results of the
experiments with neutral current detection processes should be
recalculated for the case of the charged current.

In the approach under consideration, formula
(\ref{PeeM_adiab_asymp}) with $\big( {K_{ei}^{(\text{f})} }
\big)^2 = \left| {U_{ei} } \right|^2$ provides a good
approximation for the experimentally observed electron neutrino
survival probability, since the normalized differential neutrino
detection probability is defined by the distance-dependent
neutrino propagator in matter and is the same for all neutrino
production processes. Numerical estimates of this function give
the result, which is in a very good agreement with the result
presented in Fig. 3 of paper \cite {Borexino}.

The mean value of the process probability per unit time that would
be  experimentally observed on Earth is given by formula
\eqref{dif_W_as}, where
\begin{equation}
\frac{{\text{d} W_{\text{p},i} \left( E \right) }}{\text{d} E} = \big(
{K_{ei}^{(\text{i})} } \big)^2 \, \frac{{\text{d} W_{\text{p}} \left( E
\right) }}{\text{d} E}
\end{equation}
is now the production probability per unit time  of the neutrino
mass eigenstates in the Sun and the cross section of the neutrino
mass eigenstate detection in vacuum $\sigma_{\text{d},i} \left( E
\right)$ is given by formula (\ref{cross_sect_vacuum}).

As we mentioned at the end of Section 2, such a factorization of
the process probability takes place for all solar neutrino
production processes and, therefore, the sum of the production
probabilities of all these processes, which is in a good agreement
with the experimental data, is expressed as the sum of the
production and detection probabilities of neutrino mass
eigenstates, without taking oscillations into account.

\section{Conclusion}
The solar neutrino problem  arose in the late 1960s as a
discrepancy between the predicted neutrino flux and the lower
observed number of detected neutrinos \cite{Bahcall:1972}. It was
in the epoch of flavor states of massless neutrinos, and the
numbers of events in detectors were calculated with the formulas
similar to (\ref{dif_W_el}).

In the 1970s a solution to the solar neutrino problem was put
forward based on the notion of flavor states of massive neutrinos
\cite{Bilenky:1976cw,Bilenky:1976yj}, which explained the neutrino
deficit with the help of neutrino oscillations giving formulas for
the number of events of the type (\ref{dif_W_tot}). This approach,
complemented by the Mikheev-Smirnov-Wolfenstein mechanism \cite{Mikheyev:1985zog,Mikheev:1986wj}, is now
standard and seems to be capable of describing all neutrino
interaction processes at large distances. However, from the point
of view of quantum mechanics, these flavor states are ill-defined
and cannot be consistently described within quantum field theory
\cite{Blasone:2019rxl}.

In the present paper, we took the next step and showed that all
results of the standard quantum-mechanical approach can be
reproduced within a quantum field-theoretical description without
referring to neutrino flavor states. In particular, we developed a
quantum field-theoretical description of neutrino oscillations in
matter in the adiabatic approximation. This approach yields the
correct numbers of neutrinos that would be detected in experiments
on Earth, if one counts the number of produced and detected
neutrino mass eigenstates, which is given by formula
(\ref{dif_W_as}), rather than the number of neutrino flavor
states. Thus, neutrino oscillations, which really exist and are
observed in experiments such as T2K, actually play no role in
solving the solar neutrino problem. It is solved by using only the
neutrino mass eigenstates coupled to the charged leptons via the
PMNS matrix. It is worth noting that dispensing with neutrino
flavor states also restores quark-lepton symmetry, since flavor
states are absent in the quark sector.

Finally, the proposed approach correctly reproduces the dependence
of the normalized differential neutrino detection probability
(called the electron neutrino survival probability in the standard
quantum-mechanical description) on the neutrino energy.

These results unambiguously indicate that the true physical states
of neutrinos are the mass eigenstates, whereas the flavor states,
which are a relic of the massless neutrino epoch, can serve only
as auxiliary states in calculations using the approximation of
zero neutrino masses.

\section*{Acknowledgments}
The authors are grateful to E. Boos, \fbox{A. Lobanov,} M. Libanov
A. Pukhov, and  \fbox{V. Rubakov.} for interesting and useful
discussions. The authors express their special gratitude to M.
Smolyakov  for reading the manuscript and for his valuable
comments. \vspace{5mm}

\noindent The study was conducted under the state assignment of
Lomonosov State University.

\section*{Appendix: Equation for $m_{i\text{M}}^2$ and its solutions}

By making the change of variable $m_{i\text{M}}^2  = t_i  + m_2^2$
in  equation \eqref{mM2_eq} and introducing the notation $\Delta
m^2  = m_2^2  - m_1^2$ and $\Delta M^2  = m_3^2  - m_2^2$ for the
squared masses differences, we obtain the equation
\begin{equation}
\begin{split}
& t_i^3  + \left( {\Delta m^2  - \Delta M^2  - 2pf} \right)t_i^2 \\
& - \left\{ {\Delta m^2 \Delta M^2  + 2pf\left[ {\Delta m^2 \left(
{1 - \left| {U_{e1} } \right|^2 } \right) - \Delta M^2 \left( {1 -
\left| {U_{e3} } \right|^2 } \right)} \right]} \right\}t_i  +
2pf\Delta m^2 \Delta M^2 \left| {U_{e2} } \right|^2 = 0.
\end{split}
\end{equation}
Making another change of variable to bring this equation to the
canonical form, solving it using Cardano's formula and requiring
$m_{i\text{M}}^2 = m_i^2$ for $f^\mu   = 0$, we get the
eigenvalues in the form
\begin{align}
& m_{1\text{ M}}^2  = m_2^2  - \frac{1}{2}\left( {\alpha  + \beta } \right) - \frac{1}{3}\left( {\Delta m^2  - \Delta M^2  - 2pf} \right) + \text{i}\frac{{\sqrt 3 }}{2}\left( {\alpha  - \beta } \right), \\
& m_{2\text{ M}}^2  = m_2^2  - \frac{1}{2}\left( {\alpha  + \beta } \right) - \frac{1}{3}\left( {\Delta m^2  - \Delta M^2  - 2pf} \right) - \text{i}\frac{{\sqrt 3 }}{2}\left( {\alpha  - \beta } \right), \\
& m_{3\text{ M}}^2  = m_2^2  + \alpha  + \beta  - \frac{1}{3}\left(
{\Delta m^2  - \Delta M^2  - 2pf} \right),
\end{align}
where
\begin{equation}
\alpha  = \sqrt[3]{{ - \frac{b}{2} + \sqrt Q }}, \qquad \beta  =
- \frac{c}{{3\alpha }}, \qquad Q = \frac{{c^3 }}{{27}} +
\frac{{b^2 }}{4},
\end{equation}
\begin{equation}
c = - \Delta m^2 \Delta M^2  - 2pf\left[ {\Delta m^2 \left( {1 -
\left| {U_{e1} } \right|^2 } \right) - \Delta M^2 \left( {1 -
\left| {U_{e3} } \right|^2 } \right)} \right] - \frac{1}{3}\left(
{\Delta m^2  - \Delta M^2  - 2pf} \right)^2 ,
\end{equation}
\begin{equation}
\begin{split}
b = & \, \frac{2}{{27}}\left( {\Delta m^2  - \Delta M^2  - 2pf} \right)^3  + 2pf\Delta m^2 \Delta M^2 \left| {U_{e2} } \right|^2  \\
& + \frac{1}{3}\left( {\Delta m^2  - \Delta M^2  - 2pf}
\right)\left\{ {\Delta m^2 \Delta M^2  + 2pf\left[ {\Delta m^2
\left( {1 - \left| {U_{e1} } \right|^2 } \right) - \Delta M^2
\left( {1 - \left| {U_{e3} } \right|^2 } \right)} \right]}
\right\}.
\end{split}
\end{equation}
Note that $m_{1\text{M}}^2  < m_{2\text{M}}^2  < m_{3\text{M}}^2$ if $m_1^2  < m_2^2  < m_3^2$.

\end{document}